\documentclass[aps,twocolumn,preprintnumbers,amsmath,amssymb,superscriptaddress]{revtex4-1}
\usepackage{epsfig}
\usepackage{color}
\usepackage{graphics}
\usepackage{graphicx}
\usepackage{amsmath}
\usepackage[caption=false]{subfig}
\begin{document}

% avoids incorrect hyphenation, added Nov/08 by SSR
\hyphenation{ALPGEN}
\hyphenation{EVTGEN}
\hyphenation{PYTHIA}

\title{Diffusion in Quasi-One Dimensional Channels: A Small System $n,p,T,$ Transition State Theory for Hopping Times.}
\author{Sheida Ahmadi}
\affiliation{Department of Chemistry, University of Saskatchewan, Saskatoon, Saskatchewan S7N 5C9, Canada}

\author{Richard K. Bowles}
\email{richard.bowles@usask.ca}
\affiliation{Department of Chemistry, University of Saskatchewan, Saskatoon, Saskatchewan S7N 5C9, Canada}

\begin{abstract}
Particles confined to a single file, in a narrow quasi-one dimensional channel, exhibit a dynamic crossover from single file diffusion to Fickian diffusion as the channel radius increases and the particles can begin to pass each other. The long time diffusion coefficient for a system in the crossover regime can be described in terms of a hopping time, which measures the time it takes for a particle to escape the cage formed by its neighbours. In this paper, we develop a transition state theory approach to the calculation of the hopping time, using the small system isobaric--isothermal ensemble to rigorously account for the volume fluctuations associated with the size of the cage. We also describe a Monte Carlo simulation scheme that can be used to calculate the free energy barrier for particle hopping. The theory and simulation method correctly predict the hopping times for a two dimensional confined ideal gas system and a system of confined hard discs over a range of channel radii, but the method breaks down for wide channels in the hard discs case, underestimating the height of the hopping barrier due to the neglect of interactions between the small system and its surroundings.
\end{abstract}

\maketitle
\section{Introduction}
% definitions
Particle motion in bulk systems is described by Fickian (normal) diffusion, where the mean square displacement (MSD) of a labeled particle increases linearly with time in the long time limit, and is given by the Einstein relation,
\begin{equation}
\langle (x_{t}-x_{0})^{2} \rangle= \langle \Delta x_{t}^{2} \rangle=2D_{x}t,
\label{eq:NormalDiffusion}
\end{equation}
where $D_{x}$ is the diffusion coefficient along an arbitrary direction $x$, $x_{t}$ is the position of the particle at time, $t$, and $x_{0}$ is the initial particle position. However, the fundamental nature of diffusion can change dramatically if the movement of the particles is restricted to a quasi-one dimensional narrow channel in which the particles cannot pass each other~\cite{Gelb:1999p1920,Gubbins:2010p11770}. The geometric restriction, when combined with the presence of independent stochastic forces~\cite{Levitt:1973p14144,Mon:2003ed} or a Brownian background~\cite{Percus:1974wp} acting on the particles, results in a form of anomalous diffusion known as single file diffusion (SFD), because the relative order of the particles is preserved and a particle must wait for its neighbors to move before being able to diffuse. This effect reduces the translational mobility of the particles in a way that causes the MSD to increase proportionally to the square root of time in the long time limit and is described by an Einstein--like relation,
\begin{equation}
\langle \Delta x_{t}^{2} \rangle=2F_{x}t^{1/2},
\label{eq:SFD}
\end{equation}
where the diffusion coefficient has been replaced by a mobility factor, $F_x$~\cite{Hahn1998,Lin:2005p15430}. 

%applications of SFD
Single file diffusion occurs in a variety of natural and engineered systems, including ion and water transport through biological membranes~\cite{Hodgkins:1955ep,Finkelstein:1987tu,Jensen:2002eh,Rasaiah:2008ef}, diffusion in molecular sieves such as zeolites~\cite{Karger1992,Karger:2015bu,Kumar:2014hf}, carbon nanotubes~\cite{Mukherjee2010}, or metal organic frameworks~\cite{Salles:2011p15709,Jobic:2016}, and charge-carrier diffusion in one dimensional channels~\cite{Kharkyanen:2010fv}. At the molecular level, pulsed field gradient nuclear magnetic resonance (NMR) techniques have been used to study SFD in adsorbate molecules confined in zeolites~\cite{Gupta1995,Kukla:1996ke, Hahn1996} and gas mixtures confined to polycrystalline dipeptide channels~\cite{Dutta:2016}. However, the observation of SFD by NMR  is subject to some uncertainty in systems where alternative diffusion mechanisms could be responsible~\cite{Das:2010,Valiullin:2010}. Single file diffusion has also been observed in colloidal systems~\cite{Wei:2000ei,Lutz:2004gr}, where it is being used in nano- and micro-fluidic devices~\cite{Siems:2012cp,Locatelli:2016kf}, as well as to control flow rates in the delivery of drugs~\cite{Yang:2010p15731}.

% crossover regime.
Fluids in quasi-one dimensional channels also exhibit an interesting dynamical crossover regime from SFD to normal diffusion as the channel radius increases beyond a passing threshold that allows the particles to hop past each other~\cite{Mon:2002p4642,Chen:2010p11420,Sane2010,Lucena:2012df,HerreraVelarde:2016dq}. In this crossover regime, the MSD in the long time limit is described by Eq.~\ref{eq:NormalDiffusion} because the particles can pass each other. However, the particles are trapped between their neighbors for long periods of time, during which they perform SFD, because passing is difficult. The resulting MSD at intermediate times is then described by Eq.~\ref{eq:SFD}. In the case of mixtures formed from particles of different sizes, it is also 
possible to select a channel radius that induces dual--mode diffusion, where the large particle component performs SFD while the smaller components exhibit normal diffusion~\cite{Ball:2009p6393,Liu:2010p15345}. The difference in particle mobilities can potentially be used for separation~\cite{Adhangale:2003p15701,Wanasundara2012}.

Mon and Percus~\cite{Mon:2002p4642} used a simple phenomenological model to show that the long time diffusion coefficient of a system in the crossover regime can be written,
\begin{equation}
D_{x} \sim 1/ \tau_{hop}^{\gamma}.
\label{eq:Diff_tauhop}
\end{equation}
where $\tau_{hop}$ is the average time for a particle to escape the cage formed by its two nearest neighbors and the exponent $\gamma$ depends on the nature of the hopping event. If the system is near the passing threshold for the particles, so that hopping events are rare and the particles are trapped long enough to perform SFD, then $\gamma$ is predicted to be $1/2$, and this has been confirmed by simulation for quasi-one dimensional hard discs~\cite{Mon:2006p2267,Mon:2007p2259}, hard spheres~\cite{Mon:2002p4642} and soft-potential systems such as Lennard--Jones~\cite{Wanasundara:2014hw}, as well as for mixtures~\cite{Ball:2009p6393}. The hopping time approach is useful because the important details that influence the long time dynamics of the particles, such as the particle--particle and particle--wall interactions, as well as the density or pressure, are captured in a single, short time, phenomenological parameter that is also accessible by simulation and theory.

%However, when the channel radius is wide enough for the particles to pass easily, it is expected that the hopping time will be short and the distance that a particle travels between hops will be independent of $\tau_{hop}$ and proportional to the average distance between nearest neighbors in the channel, which leads to the exponent $\gamma=1$. 

% papers goals
The process by which particles pass each other in a narrow quasi-one dimensional system can be treated in terms of a free energy barrier crossing process because the particle--particle and particle--wall interactions combine to increase the energy of interaction, and restrict the available configuration space for two particles, as they attempt to pass.  This suggests that transition state theory (TST) could provide a useful description of the hopping time. In the case of hard sphere systems, $\tau_{hop}$ diverges as the radius of the channel decreases towards the passing threshold from above because the infinite nature of the interaction potential eventually prevents the particles from ever passing and TST correctly predicts the exponent for the power law behavior~\cite{Bowles:2004p7}. Transition state theory also provides a qualitative description of $\tau_{hop}$ for soft potential systems, where the passing threshold is not as clearly defined~\cite{Wanasundara:2014hw}. However, these approaches are not quantitative. It is also worth noting that TST has been used to study diffusion in zeolite systems where the diffusing particles move between large structural cages formed in the channel by hoping through narrow windows~\cite{Chmelik:2016bj}. In these systems, the cage volumes remained fixed and traditional TST is effective. However, in the case of structureless channels, such as the ones considered here, the particles are trapped by their neighbours so that the cages are dynamic and have fluctuating volumes. The goal of this paper is to develop a rigorous approach to the calculation of the hopping time, within the framework of transition state theory, that can provide the basis for the quantitative prediction of $\tau_{hop}$. To achieve this, we develop the theory and associated simulation method using the small system isobaric--isothermal ensemble to account for the fluctuating nature of the nearest neighbour cage that traps a particle in a single file fluid.

% outline
The remainder of the paper is organized as follows: Section~\ref{sec:bt} provides a brief summary of the background and general derivation of the small system $n,p,T$ ensemble in order to highlight the key elements of the method. This is followed by the description of how the small system $n,p,T$ ensemble can be used to develop a transition state theory for hopping times in confined single file fluids. Section~\ref{sec:sim} outlines the Monte Carlo (MC) simulation methods used to calculate the transition state free energy barriers associated with particle hopping and the direct measurement of the hopping times in a large system. Section~\ref{sec:apps} describes the application of the theory and simulations for two simple confined systems: a two dimensional system of ideal gas particles confined between hard walls and a two dimensional system of hard discs confined between hard walls. {We have chosen to focus on these two dimensional systems because they are analytically tractable, which allows us to check our simulation method, and our results can be compared to earlier hopping time studies performed in two dimensions. However, the two dimensional geometry is also relevant to the dynamics of colloidal particles confined to lithographically formed channels~\cite{Wei:2000ei}, where the particles are effectively restricted to two dimensions by the gravitational potential well within the channel. Section~\ref{sec:dis} contains our discussion and our conclusions are summarized in Section~\ref{sec:con}.

\section{Background and Theory}
\label{sec:bt}
\subsection{The Small System $n,p,T$ Ensemble}
% general description of NPT.
In the thermodynamic limit, fluctuations away from the most probable microstates are small and the thermodynamics of the system can be described by the properties of the states associated with the maximum term in the probability distribution. The maximum term for one ensemble can be shown to be degenerate with the partition function of another so that all the ensembles are equivalent and selecting a particular ensemble for the calculation of a system property becomes a matter of mathematical convenience. However, as the system size decreases, fluctuations become more important and in the nano-scale regime it is essential to consider the types of fluctuations available to the system in selecting an appropriate ensemble. In the case of the isobaric-isothermal, $(N,P,T)$ ensemble, where the volume fluctuates, it also becomes necessary to consider the details of how the ensemble itself is formulated~\cite{Brown:2010ik,Munster:1959fh,Sack:1959jq,Attard:1995bg,Koper:1996iz,Corti:1998gb,Corti:2001ib,Corti:2002ji}. The original formulation of the ensemble by Guggenhiem~\cite{Guggenheim:1939bo} can be expressed in integral form~\cite{Hill:Y6TE6auF} as
\begin{equation}
\Delta(N,p,T)=(V_o)^{-1}\int_V Q(N,V,T)e^{-pV/kT}\mbox{,}\\
\label{eq:NPTO}
\end{equation}
where $Q(N,V,T)$ is the canonical partition function for $N$ particles contained in a volume $V$ at temperature $T$, $k$ is Boltzmann's constant and $p$ is the external pressure being applied to the system. The prefactor contains a volume scale, $V_o$, to ensure the partition function remains dimensionless. In the thermodynamic limit the choice of $V_o$ is arbitrary and it is usually taken to be $kT/p$, but for small systems, the value of the volume scale depends on the properties of the system such as the number of particles and the details of the boundary defining the volume.

% Corti small system npT.
To address these concerns Corti \cite{Corti:2001ib} developed a rigorous approach to the isobaric-isothermal ensemble for small systems that avoids the need to specify a volume scale. Here, we will provide a brief summary of Corti's derivation for a small system, $n,p,T$ ensemble, highlighting the key elements of the method and how they pertain to our particular application.
The analysis begins by considering  the division of the canonical partition function for $N$ particles, in a volume $V$ and at a temperature $T$, into a small subsystem consisting of $n$ particles in a sub--volume $v$ that is located at a point $r_0$ and surrounded by a system of $N-n$ particles in the remaining volume, $V-v$ (see Fig.~\ref{fig:shell}(a)). Separating the subsystem from its surroundings is a  mathematical boundary that has no mass or momentum and is assumed to be spherical. To avoid an over counting of configurations caused by the fluctuation of the volume that does not involve a change in particle coordinates, it is necessary to tie the location of the boundary to a shell molecule. This involves requiring that at least one of the $n$ particles of the subsystem is located in a shell, $dv$, at the boundary. Now when the volume fluctuates, the shell molecule moves with the boundary to ensure the system samples a new configuration.

%General figure for n,p,T ensemble.
\begin{figure}[h]
\centering
\includegraphics[width=2.8in]{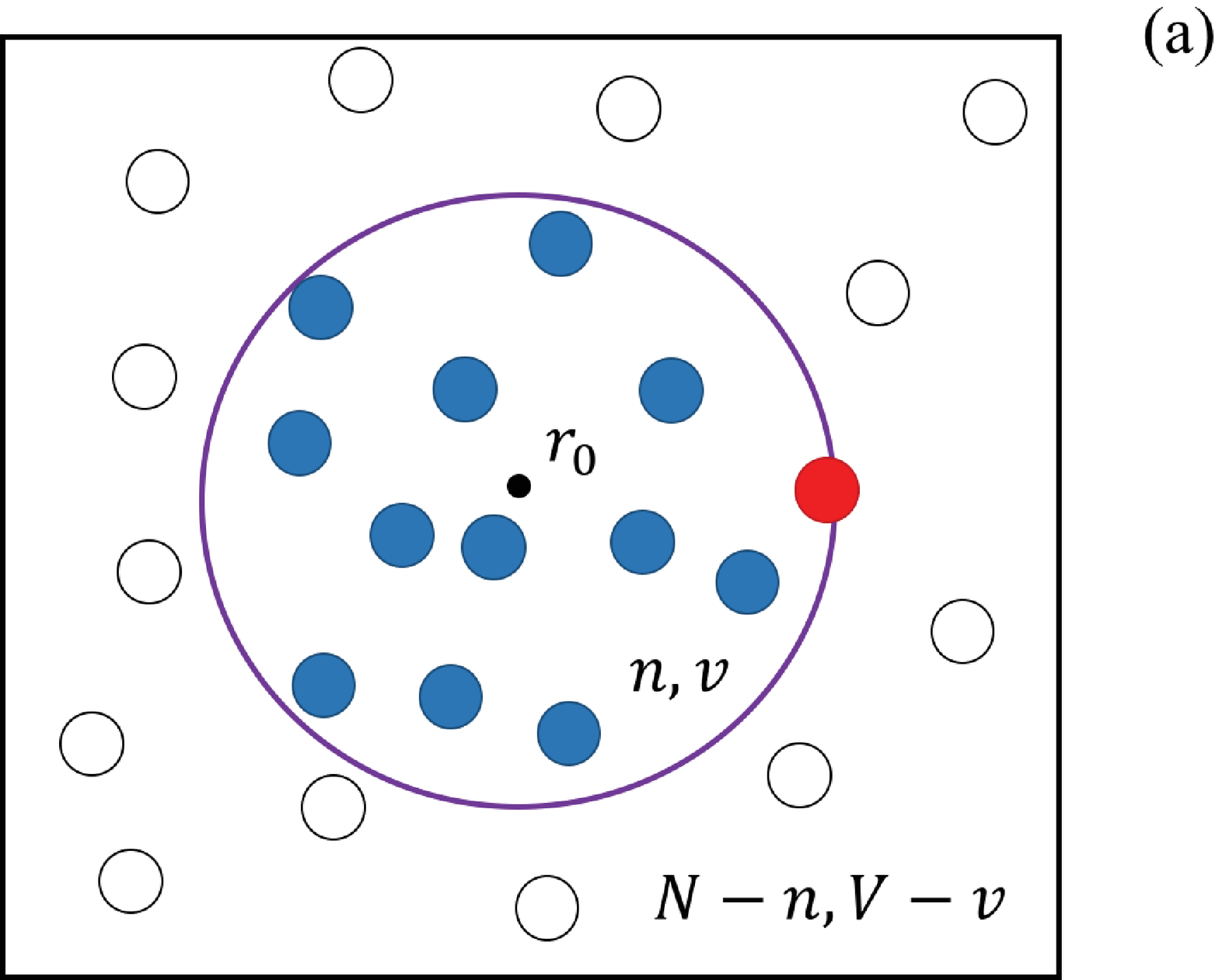}
\includegraphics[width=2.8in]{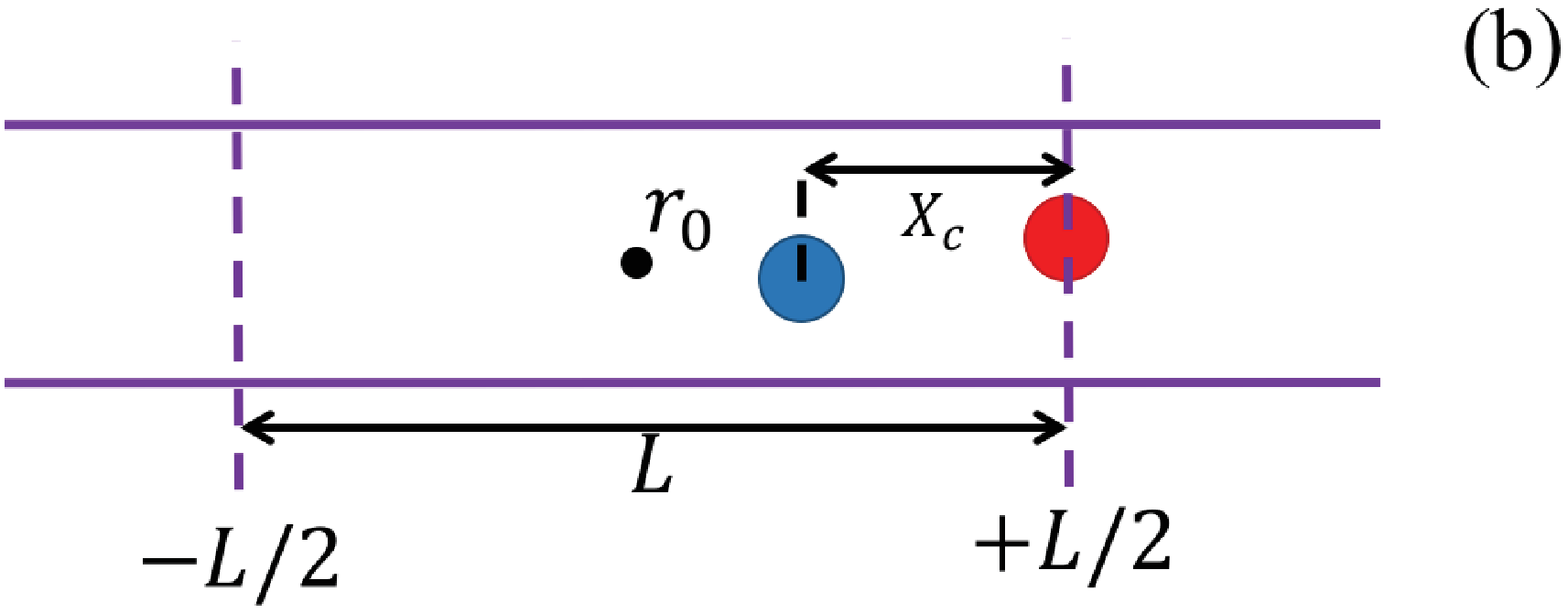}
\includegraphics[width=2.8in]{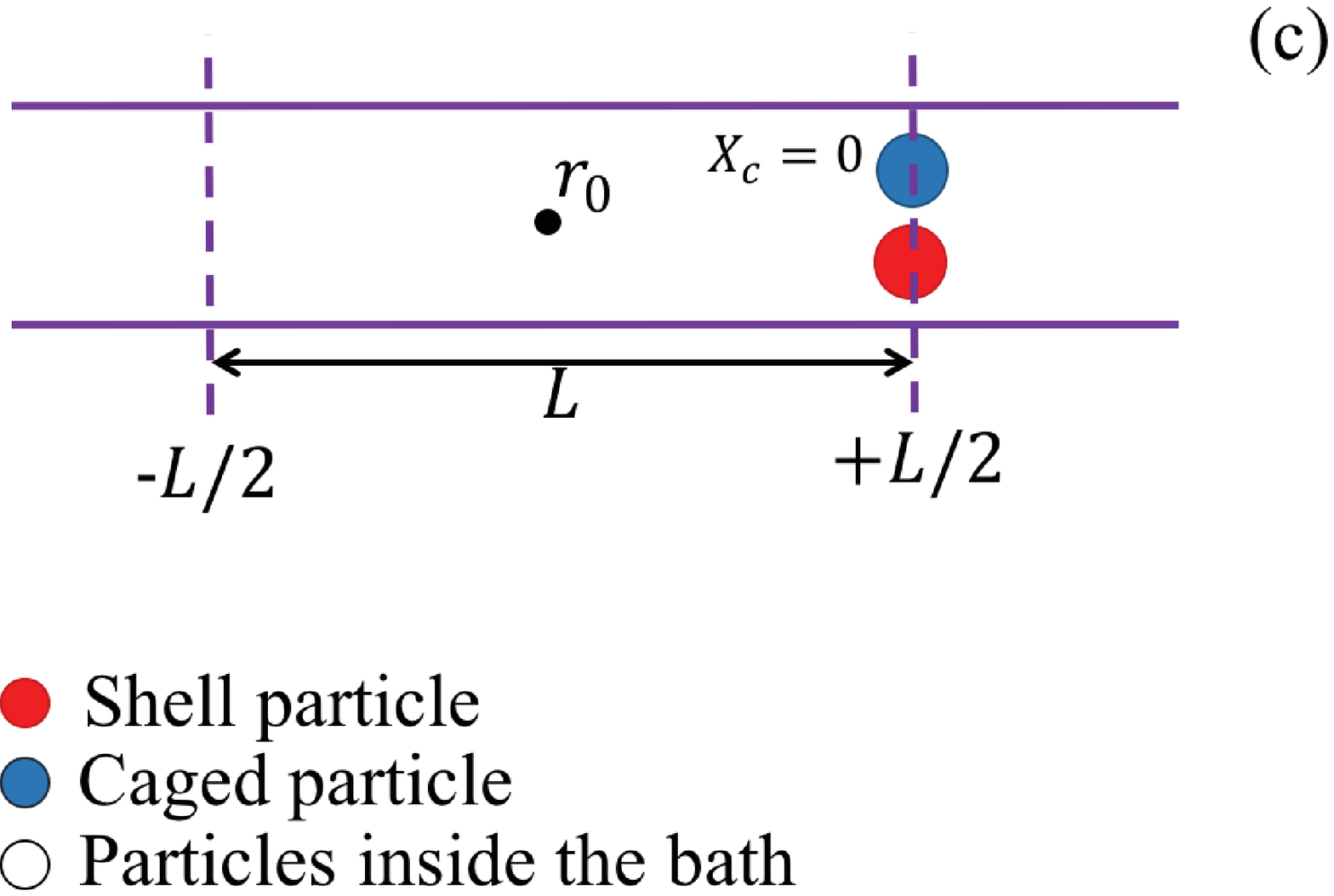}
\caption{(a) A configuration of a bulk canonical system decomposed into a small $n,v,T$-subsystem, located at $r_0$, with a shell molecule, and its $N-n,V-v,T$ surroundings. (b) The two dimensional, quasi-one-dimensional small $n,p,T$ system, located at $r_0$, with a shell particle located at $L/2$ and the caged particle located at a distance along the axial direction, $x_c$, relative to the shell particle. (c) The transition state for the hopping process with $x_c=0$.}
\label{fig:shell}
\end{figure}

A canonical partition function for the $n$ particles in the volume $v+dv$ can be written as,
\begin{equation}
Q_{n,v}^{*}dv=\frac{dv}{(n-1)! \Lambda^{Dn}} \int_{v+dv} e^{-\beta (U_{n}+U_{\sigma})}(dr)^{n-1}\mbox{,}\\
\label{eq:Delta1}
\end{equation}
where $\Lambda$ is the de Broglie wavelength, $D$ is the dimensionality, $U_{n}$ is the potential energy of the $n$ particles in the small volume interacting amongst themselves, $U_\sigma$ is the interaction between the particles in the subsystem and the $N -n$ particles in the surroundings and 
$(dr)^{n-1}=dr_{1,2}\cdots dr_{1,n}$ are the particle volume elements with positions relative to the location of the shell particle. Furthermore, the integration of the shell particle, over the volume in the shell, $dv$, and the other $n-1$ particles over $v+dv$, is performed with the positions of the surrounding $N-n$ particles held fixed so the canonical partition function for the complete $N,V,T$ system can be expressed,
\begin{equation}
Q(N,V,T)=\int_0^{V}Q(N-n,V-v,T)\left<Q_{n,v}^{*}\right>_0 dv\mbox{,}\\
\label{eq:qnvt}
\end{equation}
where, 
\begin{equation}
\left<Q_{n,v}^{*}\right>_0=\frac{\int_{V-v}Q_{n,v}^{*}e^{-\beta U_{N-n}}(dr)^{N-n}}{\int_{V-v}e^{-\beta U_{N-n}}(dr)^{N-n}}\mbox{,}\\
\label{eq:qave}
\end{equation}
is an average over the configurations of the surrounding $N-n$ particles interacting amongst themselves with a potential energy $U_{N-n}$.

Now, the probability of finding the system with $n$ particles in a volume $v+dv$ and the remaining particles outside can be written,
\begin{equation}
\begin{split}
P_n(v)dv&=\frac{Q(N-n,V-v,T)\left<Q_{n,v}^{*}\right>_0 dv}{Q(N,V,T)}\\
&=e^{\beta n \mu}e^{-\beta W(v)}\left<Q_{n,v}^{*}\right>_0 dv\mbox{,}
\label{eq:pdv}
\end{split}
\end{equation}
where $\mu$ is the chemical potential of the surroundings, $W(v)$ is the work required to form an empty cavity in the system and the following relation has been used:
\begin{equation}
\frac{Q(N-n,V-v,T)}{Q(N,V,T)}=e^{\beta n \mu}e^{-\beta W(v)}\mbox{.}\\
\label{eq:mu1}
\end{equation}

The small system $n,p,T$ partition function can then be written,
\begin{equation}
\Delta= \int \left<Q_{n,v}^{*}\right>_0 e^{-\beta W(v)} dv\mbox{,}
\label{eq:Delta0}
\end{equation}
by using the normalization condition, $\int_v P_n(v)dv=1$ and noting $\Delta=\exp(-\beta n\mu)$. Depending on the size and shape of the cavity, $W(v)$ may contain 
work in addition to the usual volume work, such as surface work, so that the pressure inside the system, $\hat{p}$, does not necessarily equal the imposed pressure. However, if the interface is planar, or the small subsystem becomes macroscopic, then $\hat{p}=p$  and $W(v)=pv$. If the interactions between the subsystem and its surroundings are also negligible, Eq.~\ref{eq:Delta0} becomes,
\begin{equation}
\Delta= \int Q_{n,v}^{*} e^{-\beta pv} dv\mbox{.}
\label{eq:Delta1}
\end{equation}

Most notably, Eq.~\ref{eq:Delta1} contains no volume scale because the uncertainty associated with identifying new configurations in phase space as the volume of the container fluctuates is avoided by the use of the shell molecule. However, considerable care needs to be taken when implementing the new small system $N,p,T$. For example, the integral over the shell particle position performed to obtain Eq.~\ref{eq:Delta1} relies on the underlying spherical symmetry of the system volume. The same approach can be taken for systems with alternative shapes such as cubes etc, but this is not always possible when considering non-isotropic systems, as we will do in this work, and alternative approaches to evaluating degrees of freedom associated with the shell particle may be needed.

Corti~\cite{Corti:2002ji} also showed that the small system isobaric-isothermal partition function is related to the thermodynamic potential,
\begin{equation}
G=\left<U^*\right>+p\left<v\right>-TS\mbox{,}
\label{eq:ghat1}
\end{equation}
where $\left< \cdots\right>$ denotes the ensemble average and $U^*$ is the internal energy in the presence of a shell particle, through the expression
\begin{equation}
G=-kT\ln\Delta\mbox{.}\\
\label{eq:ghat2}
\end{equation}
It is important to note that $G$ does not necessarily equal the Gibbs free energy of the {\it system}, except in the thermodynamic limit, because it contains information regarding the surroundings, as is evident from Eq.~\ref{eq:pdv} where the chemical potential of the surrounding is introduced, and because the small system samples a different set of volume fluctuations compared to a thermodynamically sized system. This is the case even when $U_{\sigma}=0$, and there is no interaction between the system and the surroundings. Consequently, a change in $G$, defined by Eq.~\ref{eq:ghat2}, represents the reversible work of moving the system between two states, including contributions from the surroundings. A more complete discussion of these subtleties can be found in references \cite{Corti:1998gb,Corti:2001ib,Corti:2002ji}.

% adaptation to our system
\subsection{A Transition State Theory for Hopping Times}

%brief description of Transition state theory. (need good general reference for TST)
Transition state theory is an equilibrium based approach to the calculation of transition rates in activated processes, originally developed by Eyring~\cite{Eyring:1935dt} to describe rates in chemical reactions. While the approach neglects the effect of barrier recrossing, it provides an upper bound to the rate and more sophisticated methods, such as those proposed by Bennet~\cite{Bennett:1975ut} and Chandler~\cite{Chandler:1978hh}, reduce to TST in the regime where the barrier is high relative to the thermal energy in the system and recrossing is rare.

According to TST, the time it takes for a barrier crossing event like hopping to occur is inversely proportional to the rate and is given by,
\begin{equation}
\tau_{hop} =\frac{\kappa}{P^*}=\kappa e^{\beta \Delta G^*}\mbox{,}\\
\label{eq:t_hop_TS2}
\end{equation}
where $\kappa$ is a kinetic prefactor, $P^*$ is the probability of finding the system in the transition state and $\Delta G^*$ is the height of the free energy barrier for the activated process.

Figure~\ref{fig:shell}(b) shows the construction of the small $n,p,T$ system used here to study the TST free energy barrier associated with a particle escaping its cage in a quasi-one dimensional channel by hopping past one of its neighboring particles. We consider an $n=2$ particle system confined to a long, uniform, two dimensional (2D) channel that extends longitudinally along the $x$-axis and has a radius $R_p$ in the $y$-axis. The cell center is located at a point $r_0$ and the shell particle is located at $+L/2$ so that the total length of the cell is $L$ and its two dimensional volume is $v=2R_p L$. The channel radius remains fixed so that fluctuations in the volume are given by,
\begin{equation}
dv=2R_pdL\mbox{.}\\
\label{eq:dv}
\end{equation}
The small system isobaric--isothermal partition function given by Eq.~\ref{eq:Delta1} then becomes,
\begin{equation}
\Delta= \int Q_{n,v}^{*} e^{-\beta p_l 2R_p L } 2R_p dL\mbox{,}
\label{eq:Delta2}
\end{equation}
where $p_l$ is the longitudinal pressure acting on the end of the channel~\cite{KOFKE:1993p3863}. Particle 1 is the shell particle placed on the boundary at one end and the $2R_p$ factor in $dv$ arises from the integration of the shell particle over its $y$-axis degree of freedom.

The second particle in our system represents the caged particle. If we define a reaction coordinate for the hopping process as the axial separation of the caged particle from the shell particle, $x_c=x_1-x_2$, then the transition state occurs at $x_c=0$ as shown in Fig.~\ref{fig:shell}(c). The configuration space associated with specific points along the reaction coordinate can be measured by applying a {\bf delta function, $\delta (x_{c}^{\prime}-x_{c})dx_c$, that is one when $x_{c}^{\prime}-x_{c}=0$ and zero otherwise}, to Eq.~\ref{eq:Delta2}, to obtain a reaction coordinate partition function,
\begin{equation}
\Delta_{x_{c}^{\prime}} \; dx_{c}= \int_{L} Q_{n,v,T}^{*} \; e^{-\beta p_l 2R_p L } \; \delta (x_{c}^{\prime}-x_{c}) 2R_p\;dLdx_c\mbox{,}\\
\label{eq:deltaxc0}
\end{equation}
which satisfies,
\begin{equation}
\Delta= \int_{x_{c}} \Delta_{x_{c}} \; dx_{c}\mbox{.}
\label{eq:deltaxc1}
\end{equation}
The probability of having the caged particle at $x_{c}^{\prime}$ is then,
\begin{equation}
P(x_{c}^{\prime}) \; dx_{c}=\frac{\Delta_{x_{c}^{\prime}} \; dx_{c}}{\Delta}\mbox{,}\\
\label{eq:pxc}
\end{equation}
where $P(x_{c}^{\prime})$ is  a probability density and 
\begin{equation}
\int_0^{\infty} P(x_{c}) \; dx_{c}=1\mbox{.}\\
\label{eq:intpxc}
\end{equation}
Finally, the Gibbs free energy barrier for particle hopping, $\Delta G^*$, is directly related to the probability of finding the caged particle in the transition state by,
\begin{equation}
\beta\Delta G^*=-\ln P^*
\label{eq:dg0}
\end{equation}
where
\begin{equation} 
P^*=\int_0^{x_c^{*}}P(x_c)dx_c\mbox{,}\\
\label{eq:pstar}
\end{equation}
and the range of the reaction coordinate, $x_c=0$ to $x_c^*$, is the transition state region. Defined in this way, $\Delta G^*$ represents the work required to bring the caged particle into the transition state from anywhere in the cage and, as such, it will always be positive because the configuration space of the transition state is a restricted subset of all the possible configurations available to the system. Identifying the entire configuration space of the ``reactive region" as the appropriate reference state leads to a significant improvement in the predicted rates using TST in activated process as shown in the case of heterogeneous nucleation~\cite{Scheifele:2013uo}.

 The boundary separating two neighbouring cages along the reaction coordinate occurs at $x_c=0$. However, it is necessary to define a transition region with $x_c^*$ to ensure the probability $P^*$ in Eq. 21 is dimensionless, rather than a probability density. We do not have a general method for selecting $x_c^*$, but  we would expect the transition region to be small relative to size of the fundamental particle motion responsible for taking the particle from one cage to the other. If $x_c^*$ is chosen to be too large, then the transition state will contain contributions from configurations that have little or no chance of crossing the barrier in the small time limit, which would lead to poor estimates of the hopping time.

It is important to note that we have chosen to focus on a system of just two particles so that we have a single transition state to consider when a particle is attempting to hop. It could be argued that the cage of the central particle is actually formed by two particles, one on either side, giving rise to two transition states. However, each transition state involves two particles, and both of these particles escape their cages for each hopping event, which suggests we only need to consider one transition per caged particle. Explicitly including the third particle would also complicate the analysis as it would require the use of two shell molecules to define the volume. The reaction coordinate, which measures the distance of the cage particle from the transition state, would also take on two values for each configuration due to the symmetry of the system. Furthermore, the configurations of the particles outside the cell are formally included in the analysis when the full canonical partition function is divided into the small system to formal the small system $n,p,T$ ensemble. This include configurations where a particle from the surroundings sits on the boundary of the cell at $-L/2$, to form the other end of the cage.

% simulation studies. Free energy barriers and hopping time calculations.
\section{Simulation Methods}
\label{sec:sim}
\subsection{Free Energy Barrier Calculations}
Corti~\cite{Corti:2002ji} showed that the probability for accepting a trial MC move in the small system $n,p,T$ ensemble depends on the shape of the small system volume, which also constitutes the simulation cell. If we consider the case of $n$ particles confined to the two dimensional channel described in Fig.~\ref{fig:shell}(b), and ignore interactions between the particles in the cell and those in the surroundings, then the partition function for the system can be written,
\begin{equation}
\begin{split}
\Delta=&\frac{1}{(N-1)!\Lambda^{2n}}\times\\
&\int\int\int\int e^{-\beta U_n}e^{-\beta p_l 2R_p L} dLdy_1dx^{n-1}dy^{n-1}\mbox{,}
\end{split}
\label{eq:deltaMC}
\end{equation}
where the integrals over the $dxdy$ coordinates of the $n-1$ particles within the cell represent $Q^*_{n,v,T}$, the integral over $dy_1$ represents the degrees of freedom of the shell particle over the diameter of the channel and the integral over $dL$ is the degrees of freedom of the shell particle associated with the volume of the cell. If a scale factor $x_i=L\bar{x}_i$ is introduced, then Eq~\ref{eq:deltaMC} can be expressed as,
\begin{equation}
\begin{split}
\Delta=&\frac{1}{(N-1)!\Lambda^{2n}}\times\\
&\int\int\int\int L^{n-1}e^{-\beta U_n}e^{-\beta p_l 2R_p L} dLdy_1d\bar{x}^{n-1}dy^{n-1}\mbox{,}
\end{split}
\label{eq:deltaMC2}
\end{equation}
where the $U_n$ is now a function of the scaled coordinate system. Note there is no need to rescale the $y$-coordinates because fluctuations in the volume only involve changes in the longitudinal dimension of the cell. The probability of finding the system in a configuration specified by $\bar{x}^n,y^n$, in a volume with the cell length in the range $L$ to $L+dL$, is then,
\begin{equation}
P(\bar{x}^n,y^n;L)dL\propto L^{n-1}e^{-\beta U_n}e^{-\beta p_l 2R_p L}dL\mbox{,}\\
\label{eq:pmc}
\end{equation}
and the MC acceptance probability for a trial move from an old to a new configuration becomes,
\begin{equation}
\begin{split}
\mbox{acc}&\mbox{(old}\rightarrow\mbox{new)}=\mbox{min} (1,\exp \left\{\right. -\beta[U_n^{\mbox{\small{new}}}-U_n^{\mbox{\small{old}}}]\\
&-\beta p_l 2R_p[L^{\mbox{\small{new}}}-L^{\mbox{\small{old}}}]+(n-1)\ln[L^{\mbox{\small{new}}}/L^{\mbox{\small{old}}}] \left.\right\} ) \mbox{.}
\end{split}
\label{eq:mcacc}
\end{equation}
While our analysis is specific to the two dimensional channels studied here, extending Eq.~\ref{eq:mcacc} to the study of three dimensional, quasi-one dimensional channels, where fluctuations in the volume only involve changes in $L$, simply requires the appropriate geometric factor for the channel cross section to be included in the pressure-volume term.

The probability, $P(x_c)dx$, of finding the caged particle in a volume $dx$ at a distance $x_c$ from the transition state can now be calculated directly using MC simulation. The small system $n,p,T$ simulations are carried out with $n=2$ particles confined to a channel with radius, $R_p$, and length, $L$. In the case of the hard disc model, the unit of length is taken relative to the diameter of a particle $\sigma=1$. Particle one is the shell particle that defines the sub-volume and is located at $L/2$ and the center of the sub-volume is located at the origin. Particle two is placed within the cell between $-L/2$ and $L/2$. Since the two models being studied are the two dimensional ideal gas and two dimensional hard discs, the MC moves proceed as follows: A particle is selected at random and moved randomly by a step  $\delta x$ and $\delta y$, up to a maximum displacement of $|\Delta x|=|\Delta y|=0.05$. The move is immediately rejected if the trial displacement causes the particle to leave the channel or overlap with with another particle. If particle one was selected, the trial move also corresponds to a change in $L$ of $2\delta x$ and the position of particle two is rescaled to ensure it remains within the simulation cell.  Eq.~\ref{eq:mcacc} is then used to determine the probability of accepting the move. It should also be noted that the position of the shell particle must remain positive so that $V>0$.
For each state point studied, $2\times10^7$ MC moves are used to obtain equilibrium and data is collected over the next the next $10^8-10^9$ moves, depending on the height of the free energy barrier. The probability is calculated by building a histogram of configurations along the reaction coordinate using bin sizes of $\Delta x_c=0.005$ and $0.03$ for the ideal gas and hard discs respectively, and sampling the system every 1000 MC moves to ensures configurations were not correlated. Simulation runs were sufficiently long to ensure the transition state region, $x_c<\Delta x_c$, was sampled at least 100 times and the probability density was obtained by dividing the probability by the bin size.

\subsection{Hopping Time Calculations}
The hopping time, which is the average time it takes for a particle to pass one of its caging nearest neighbors in the single file, is calculated using MC simulation in the canonical ensemble. The system consists of $N=100$ particles confined to a two dimensional channel of radius $R_p$ in the $y$--direction and length $L$ along the $x$--axis, where periodic boundaries are also employed. Particles are moved using the standard Metropolis MC scheme~\cite{Frenkel2002} as a simple approximation to Brownian motion~\cite{Patti:2012uf}. A MC trial move involves the random selection of a particle, followed by the random selection of a trial step in both the $x$ and $y$--directions up to a maximum step size of $|\Delta x|=|\Delta y|=0.05$. The trial move is rejected if it causes either a particle or wall overlap and the particle is returned to its original position, otherwise it is accepted. A unit of time is defined as a single MC cycle and consists of $N$ MC trial moves. The particles are initially spaced evenly along the tube and placed randomly across the channel such that they do not overlap then the system is equilibrated over $2\times10^7$ MC cycles.  At the start of data collection, the cages for the particles are identified as their immediate right and left neighbors, and their initial hopping time is set to zero. A particle hopping time is the number of MC cycles it takes for the particle to escape its cage. After each hopping event the hopping time for the particle is recorded, then reset to zero, and the new caging neighbors are identified. Data is collected over $5\times 10^7$ MC cycles and the average hopping time, $\tau_{hop}$, is calculated over all hopping times recorded for all particles.

The hopping times for systems with different radii must be compared at the same applied external pressure. However, the calculations of $\tau_{hop}$ are carried out in the $N,V,T$ ensemble because volume fluctuations in the $N,p,T$ ensemble cause the particle positions to be rescaled, leading to possible errors. To ensure $\tau_{hop}$ is obtained at the correct state point, the density for each system with a different radius is calculated using the $N,p,T$ ensemble where $\beta P=1.0$, $N=1000$, and $10^8$ MC cycles are used after the system reaches equilibrium. A MC trial move involves the random selection of a particle, followed by the random selection of a trial step in both the $x$ and $y$--directions. The maximum step size of $|\Delta x|=|\Delta y|$ is varied  during the simulation in the range of $0.05-0.1\sigma$ to ensure $80 \%$ of the trial moves are accepted.  The shell particle approach is used. Data are sampled at every $10,000$ time steps to obtain the average volume, $\left<V\right>$, which gives the density $\rho=N/\left<V\right>$. Table~\ref{tab:densitytable} shows the range densities, $\rho=N/V$, obtained for the different radii studied in both the ideal gas and hard disc systems. The density range for hard disc two particle system is $\rho=0.54-0.65$, which is significantly higher than that obtained for the large system because the lack of interactions with the surrounding allows the system to small volume configurations that would be inaccessible if more particles were present. When a third particle is included in the small system $N,p,T$, the density range reduces to $\rho=0.47-0.53$. Nevertheless, as noted earlier, it is necessary to calculated $\tau_{hop}$ using the two particle system but this result highlights the need to ultimately account for the interactions between the cell and the surrounding, i.e. the inclusion of $U_\sigma$.
\begin{table}[ht]
\centering
\begin{tabular}{p{2.5cm}p{2.5cm}p{2.5cm}}
\hline
 Systems& $R_p$ & $\rho$ \\
\hline\hline
 2D-ideal gas  & 1.1-1.2 & 1.00 \\
2D-Hard discs & 1.01-1.2 & 0.38-0.41 \\
\hline
\end{tabular}
\caption{Selected channel radii and system densities for the idea gas and hard sphere systems at $\beta P=1.0$.}
\label{tab:densitytable}
\end{table}

% Applications
\section{Applications}
\label{sec:apps}

\subsection{Two Dimensional Ideal Gas}
To illustrate our approach, we first calculate the  hopping barrier in a two dimensional ideal gas system, where there are no interactions between particles, so that $U_{\sigma}=U_n=0$, and the hard interaction of the particles with walls keeps them confined between $-R_p$ and $R_p$. The reaction coordinate partition function for this system can be written,
\begin{equation}
\Delta_{x_{c}}dx_{c}=\frac{dx_{c}}{\Lambda^{4}}\int_{-R_p}^{+R_p}\int_{-R_p}^{+R_p}\int_{x_{c}}^{\infty}  e^{-\beta p_l 2R_p L } \; dLdy_{1}dy_{2}
\label{eq:IDG_deltaxc0}
\end{equation}
where the length of the cell is bounded in its lower limit by the position of the caged particle, and the integrals  $y_1$ and $y_2$ can be performed independently to give,
\begin{equation}
\Delta_{x_{c}}dx_{c}=\frac{2R_p}{\; \Lambda^{4} \; } \frac{e^{-\beta p_{l} 2R_px_{c}}}{\beta p_{l}} \; dx_{c}.
\label{eq:idg_deltaxc1}
\end{equation}
The partition function for the system can then be obtained by integrating over the remaining $x_c$ coordinate to yield,
\begin{equation}
\begin{split}
\Delta&=\frac{2R}{\beta p_l \Lambda^{4}} \int_{0}^{\infty} e^{-\beta p_{l} 2R_px_{c}} \; dx_{c}\\
&=\frac{1}{\beta^2 p_l^2 \Lambda^{4}}\mbox{.}
\label{eq:idg_delta1}
\end{split}
\end{equation}
To demonstrate that we have the correct partition function, we calculate the system volume,
\begin{equation}
V=2R_p\left<L\right>=-\frac{1}{\beta}\left(\frac{\partial \ln \Delta}{\partial p_l}\right)_{n,T}=\frac{2}{\beta p_l}\mbox{,}\\
\label{eq:igeos}
\end{equation}
which is the expected equation of state for an ideal gas~\cite{Corti:2001ib}.

Using Eqs.~\ref{eq:idg_deltaxc1} and \ref{eq:idg_delta1} in Eq.~\ref{eq:pxc} yields the probability of finding the caged particle at a point $x_{c}$ along the reaction coordinate as,
\begin{equation}
P(x_{c})dx_{c}=\beta p_l2R_p \; e^{-\beta p_l2R_px_{c}}\; dx_{c}\mbox{.}
\label{eq:2D_ideal_Probability}
\end{equation}
The probability of finding the caged particle within a region $dx_c$ of the transition state, where $x_c=0$, is then,
\begin{equation}
P(0)dx_c=\beta p_l2R_pdx_c\mbox{.}\\
\label{eq:p0idg}
\end{equation}
Figure~\ref{fig:idg_lnP_xc} shows that the free energy density, $-\ln P(x_c)$, increases linearly as a function of $x_c$, with a slope of $\beta p_l2R_p$, which reflects the fact that increasing $x_c$ restricts the total number of configurations available to the system by increasing the lower bound on the accessible volume fluctuations. However, it also seems to imply that the hopping process for an ideal gas particle is ``barrierless", but this is not the case. The free energy barrier $\Delta G^*$, given by Eq.~\ref{eq:dg0}, is necessarily positive because it involves an integral of the normalized probability over a small region of the reaction coordinate in the transition region. This indicates that it takes work to restrict the two particles to the transition state region relative to having the caged particle located anywhere within the small system volume. Figure~\ref{fig:idg_lnP_xc} also shows that the probability density obtained from our simulations matches the results obtained from our analytical analysis. While this is not surprising for such a simple problem, it provides a proof of principle and suggests that our simulation approach would be applicable to systems involving more complicated  particle--particle and particle--wall interactions. 

% ideal gas probability densities
\begin{figure}[ht]
\includegraphics[width=3.5in]{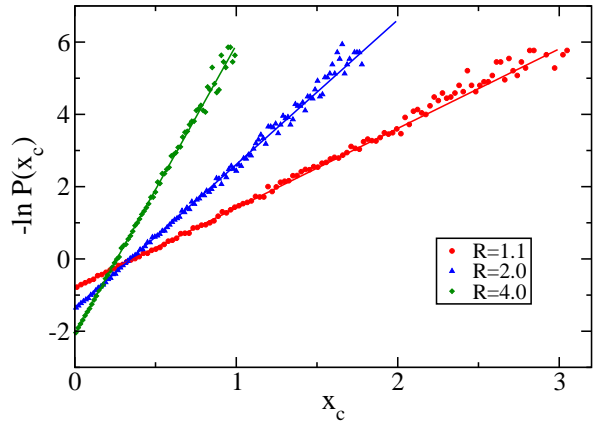}
\caption{Free energy density, $-\textup{ln} P(x_c)$ as a function of $x_c$, for three different pore radii with fixed pressure, $\beta p_l=1.0$, for the 2D ideal gas system. The solid lines represent results from the theory (see Eq.~\ref{eq:2D_ideal_Probability}). The points represent the data obtained from simulation (see Section~\ref{sec:sim} for details).}.
\label{fig:idg_lnP_xc}
\end{figure}

Figure~\ref{fig:2D_HOP_DF} shows a plot of $\ln \tau_{hop}$, obtained from our hopping time simulations versus the free energy barrier to hopping obtained from the analytical results over a range of channel radii, where we have used Eqs.~\ref{eq:2D_ideal_Probability} and \ref{eq:dg0}, with $x_c^*=0.005$. The slope is one, which suggests that our TST approach, using the small system isobaric--isothermal ensemble, provides a potentially useful method for the calculation of hopping times in these quasi-one dimensional systems. The intercept yields the kinetic prefactor $\kappa=0.90$ and essentially measures the time it takes for the particles to move out of the transition state region. Equation~\ref{eq:p0idg}, with Eq.~\ref{eq:t_hop_TS2}, suggests that $\tau_{hop}$ should vary linearly as $R_p^{-1}$, with a slope  $\sim\kappa/2\beta p_l x_c^*$ (see Fig.~\ref{fig:2D_ideal_loglog_hop_R}). This dependence on channel radius arises because our analysis is performed in the small system $n,p,T$ ensemble, where it is necessary to integrate over all possible volume fluctuations (see Eq.\ref{eq:idg_deltaxc1}). It differs from the scaling law observed for hard disks in the narrow channel regime and provides a limiting case for the hopping time in wide channels where the particles can easily pass.

% ideal gas tau a function of free energy.
\begin{figure}[ht]
\includegraphics[width=3.5in]{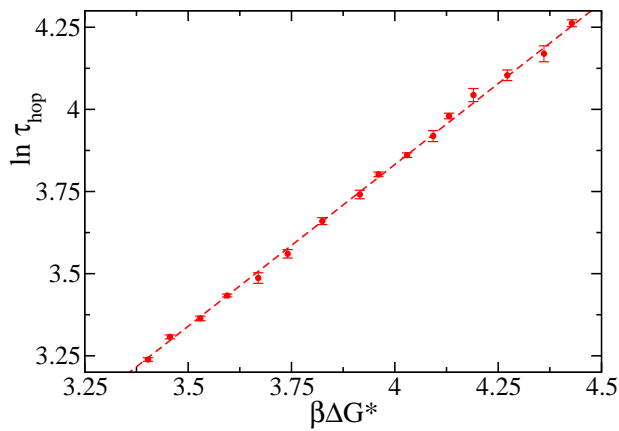}
\caption{ln$\tau_{hop}$ as a function of $\beta\Delta G^*$ for the 2D ideal gas over a range of pore radii, $R_p=1.1-4.0$. The dashed line represents the best linear fit to the data and has a slope of 0.98. The error bars represent the standard deviation in the measured hopping times.}
\label{fig:2D_HOP_DF}
\end{figure}

% ideal gas tau as a function of pore radius
\begin{figure}[ht]
\includegraphics[width=3.5in]{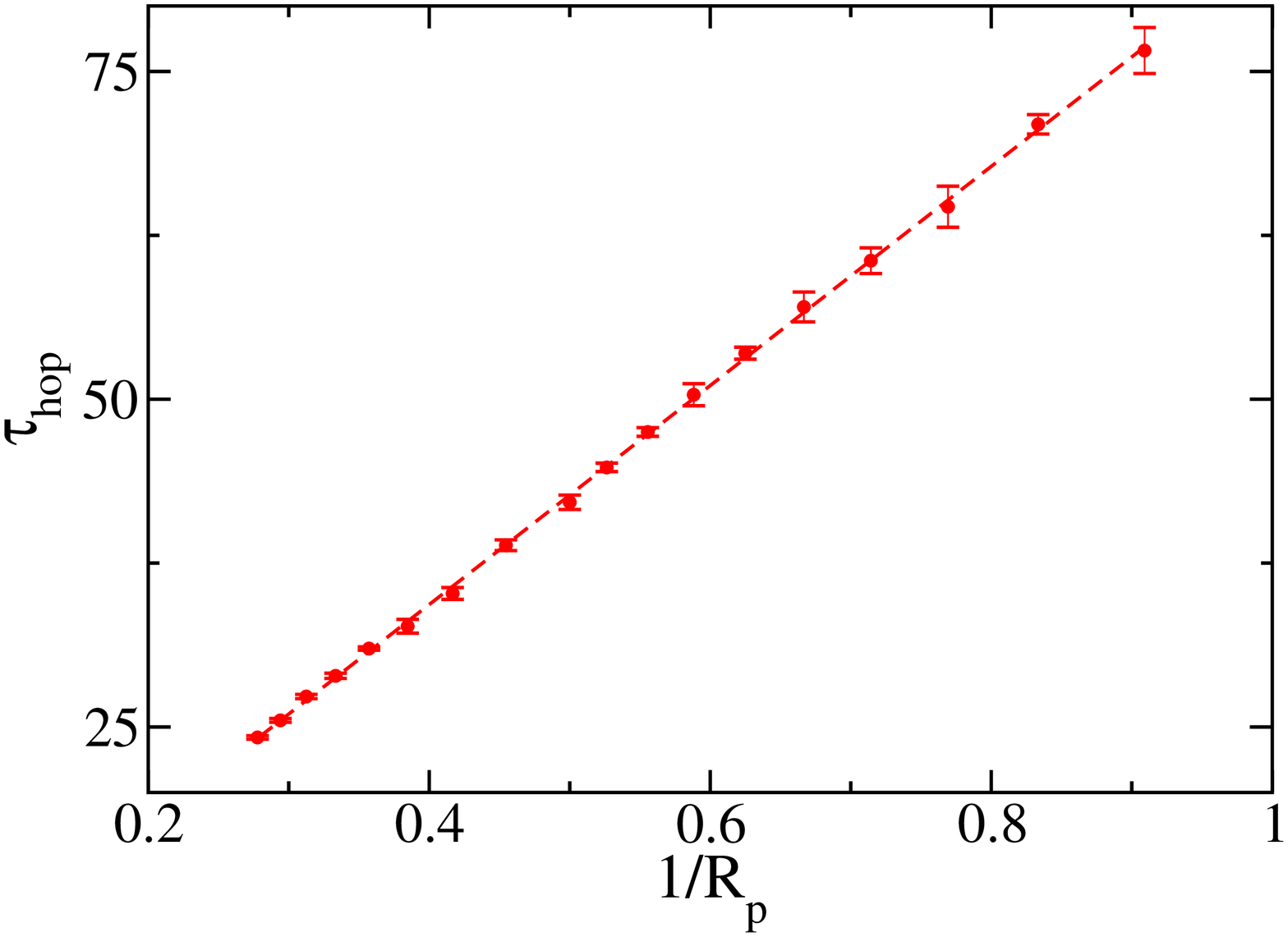}
\caption{$\tau_{hop}$, as a function of the inverse channel radius $1/R_p$, for the 2D ideal gas system. The dashed line is a linear fit to the data with a slope of 83.6. The error bars represent the standard deviation in the measured hopping times.}
\label{fig:2D_ideal_loglog_hop_R}
\end{figure}

\subsection{Two Dimensional Hard Discs}
We now calculate the hopping barrier for a system of two dimensional hard discs confined to a channel by hard walls, where the particle-particle exclusion interaction makes it more difficult for particles to escape their cage as the channel radius decreases.  For the hard disc system the particle--particle interaction is defined as,
\begin{equation}
U_{HD}(r_{ij})=
\begin{cases}
0 \ \ \textup{if} \  r_{ij} \geq \sigma, \\
\infty \ \textup{if}  \ r_{ij} < \sigma,
\end{cases}
\label{eq:HS_particle_interaction}
\end{equation}
where $\sigma$ is the particle diameter and $r_{ij}=\lvert \textup{r}_{i}-\textup{r}_{j}\rvert$ is the the distance between two particles. 
The particle--wall interaction is given by,
\begin{equation}
U_{WHS}(\hat{r}_{ij})=
\begin{cases}
0 \ \ \textup{if} \ \lvert y_{i}\rvert \leq R_p-\sigma/2, \\
\infty \ \textup{if}  \ \lvert y_{i}\rvert > R_p-\sigma/2,
\end{cases}
\label{eq:HS_wall_interaction}
\end{equation}
where $y_{i}$ is the $y$--coordinate of the particle. To make the solution to the small system $n,p,T$ partition function tractable, we consider the case where the pressure is low. The interactions between the particles inside the sub-volume and those in the surroundings are then negligible and we can assume $U_\sigma=0$.  The reaction coordinate partition function for this system is then given by,
\begin{equation}
\begin{split}
\Delta_{x_{c}}dx_{c}&=\frac{2 dx_c}{\Lambda^4}\int_{x_c}^{\infty} e^{-\beta p_l 2R_p L }dL \int_{-R_p+\frac{\sigma}{2}+\sigma_2}^{R_p-\frac{\sigma}{2}}dy_1\int_{-R_p+\frac{\sigma}{2}}^{y_1-\sigma_2}dy_2\\
&=\frac{(2R_p-\sigma-\sigma_2)^2}{\Lambda^{4}\beta p_l(2R_p-\sigma)} e^{-(2R_p-\sigma)\beta p_l\; x_c}dx_{c}\mbox{,}
\label{eq:2d_pxc}
\end{split}
\end{equation}
where,
\begin{equation}
\sigma_2=
\begin{cases}
\sqrt{\sigma^2-x_c^2} \ \ &\textup{if} \ \lvert x_c\rvert \leq \sigma,\\
\ 0 \ & \textup{if} \ \lvert x_c\rvert > \sigma\mbox{,}
\end{cases}
\label{eq:sigma2}
\end{equation}
and the factor of 2 appears because we need to include the configurations in which the order of particles 1 and 2 in the $y$-coordinate are reversed. Taking into account the piecewise nature of $\sigma_2$ and integrating over $x_c$ yields,
\begin{widetext}
\begin{equation}
\Delta=\frac{1}{\beta^2 p_l^2 \Lambda^4}\left[1+\frac{2e^{-\beta p_l(2R_p-\sigma)}\left[\beta p_l\sigma (2R_p-\sigma)+1\right]+\beta^2 p_l^2\sigma^2(2R_p-\sigma)^2-2}{\left(2R_p-\sigma\right)^4}-\frac{\pi\sigma[ I_n(z)-L_n(z)]}{2R_p-\sigma}\right]\mbox{,}\\
\label{eq:hdpf}
\end{equation}
\end{widetext}
where $I_n(z)$ is the modified Bessel function of the first kind, and $L_n(z)$ is the modified Stuve function, both with $n=1$ and $z=\beta p_l \sigma (2R_p-\sigma)$~\cite{mathfunc}.

% figire of probability density and discussion. 
Figure~\ref{fig:HS_lnP_xc} shows the free energy density, $-\ln P(x_c)$, as a function of $x_c$ for channel radii in the range $R_p/\sigma=1.01-3$. For $x_c>\sigma$, the two discs do not exclude volume with respect to each other ($\sigma_2=0$) and they effectively behave as ideal gas particles in a channel with a reduced radius of $R_p-\sigma/2$, which gives rise to the linear increase in $-\ln P(x_c)$. When $x_c<\sigma$, the two discs begin to exclude volume, leading to a restriction in configuration space. However, when the channel diameter is wide, the volume exclusion effect is small and the free energy density remains close to being linear. As the channel becomes narrower, we see the appearance of a maximum that begins at $x_c=1$ and moves towards $x_c=0$ as $R_p\rightarrow \sigma$. This unusual feature results from the convolution of two competing effects. The excluded volume interaction between the two discs reduces the number of accessible configurations available to the system as the transition state is approached. Conversely, the pressure--volume term increases the number of accessible configurations because the range of allowed volume fluctuations increases as $x_c\rightarrow 0$. The free energy density profile obtained here, in the small system isobaric-isothermal ensemble, differs significantly from that obtained in the canonical ensemble where the free energy maximum is always located at $x_c=0$~\cite{Wanasundara:2014hw}.

% Hard discs  probability densities
\begin{figure}[ht]
\includegraphics[width=3.5in]{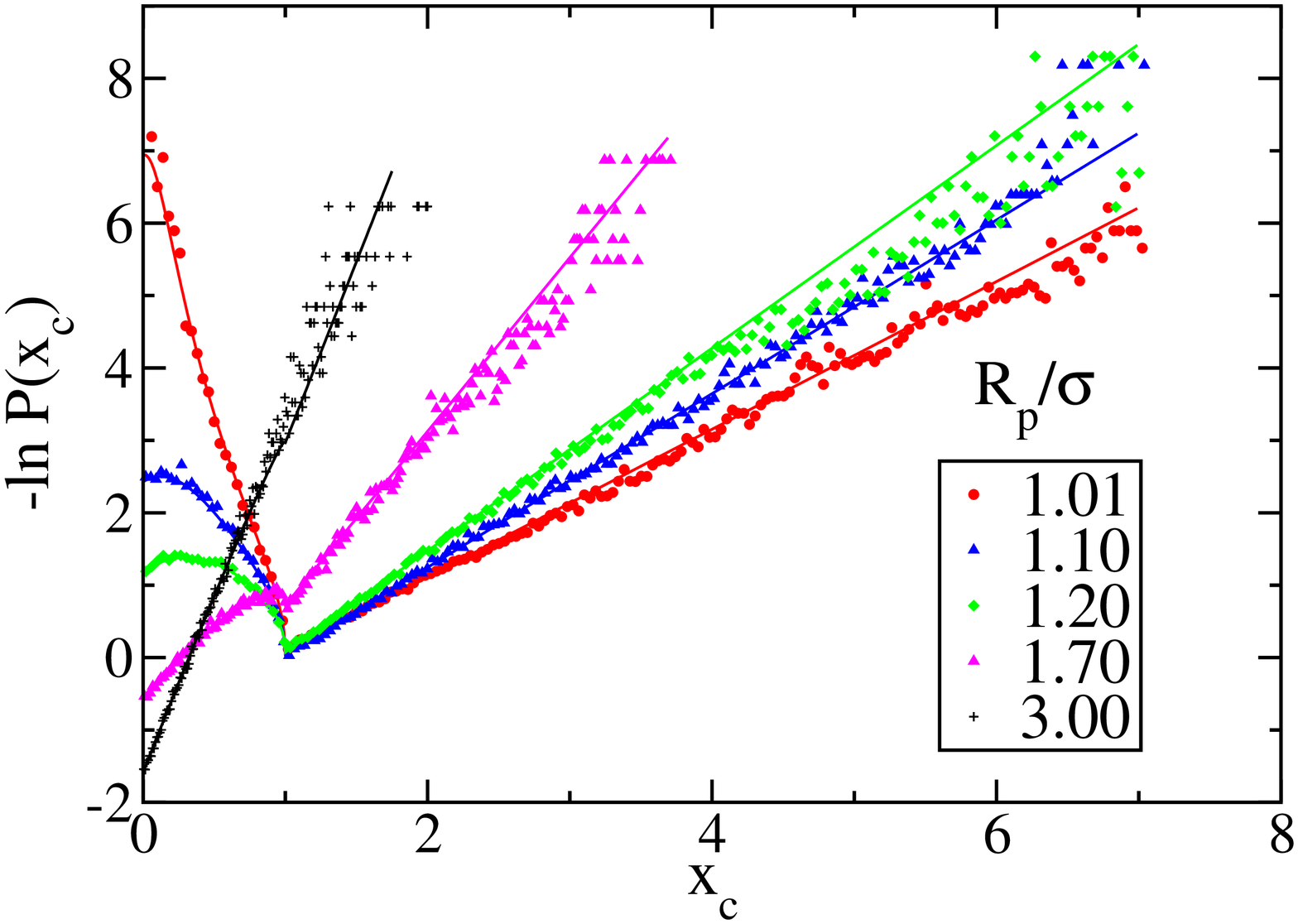}
\caption{Free energy density, $-\textup{ln} P(x_c)$ as a function of $x_c$, with different channel radii and a fixed pressure, $\beta p_l=1.0$, for the 2D hard disc system. The solid lines represent results from the theory and the points represent the data obtained from simulation.}.
\label{fig:HS_lnP_xc}
\end{figure}

In the context of particle motion, the transition state for hopping is still located at $x_c=0$, despite the presence of a maximum in the free energy density at values of $x_c>0$, because it marks the point where the particles exchange position, signifying the crossover from single--file to normal diffusion.
The probability of finding the caged particle within a region $dx_c$ of the transition state can be obtained using Eqs.~\ref{eq:2d_pxc} and \ref{eq:hdpf} in Eq.~\ref{eq:pxc}, and noting  $\sigma_2=\sigma$ (Eq.~\ref{eq:sigma2}) when $x_c=0$, to give,
\begin{equation}
P(0)dx_c=\frac{4(R_p-\sigma)^2dx_c}{\Lambda^4\beta p_l(2R_p-\sigma)\Delta}\mbox{.}\\
\label{eq:p0dxhd}
\end{equation}

The hopping time in the hard disc model is expected to diverge as a power law, $\sim(R_p-\sigma)^{\alpha}$, as the channel radius approaches the diameter of a disc because the excluded volume interaction prevents the two particles from reaching the transition state. Once the channel radius becomes smaller than the passing threshold, $R_p=\sigma$, the discs become permanently caged between their neighbours. A  Ficks--Jacobs analysis~\cite{Zwanzig:1992ig} that projects the diffusion of the two discs onto a one-dimensional reaction coordinate predicts~\cite{Kalinay:2005p2383,Kalinay:2007p2381} $\alpha=-3/2$, while TST predicts~\cite{Bowles:2004p7} $\alpha=-2$. Simulation studies of the hopping time agree with the TST result, but the scaling only becomes apparent for very narrow channels~\cite{Mon:2008wv,Mon:2007p2259}. Transition state theory suggests $\tau_{hop}\sim1/P(0)$ and Eq~\ref{eq:p0dxhd} shows that the hopping time should diverge with the exponent $\alpha=-2$, as expected. Figure~\ref{fig:HD_scaling} also shows how $1/P(0)$ crosses over from the passing threshold limit for narrow channels to the wide channel scaling behaviour $\sim(2R_p-\sigma)^{-1}$, which is consistent with our ideal gas results. Finally, Fig.~\ref{fig:HD_hop} shows that Eq.~\ref{eq:t_hop_TS2} works well for the hard disc system when the channel radius is narrow, giving rise to high barriers. This is also the regime where the hopping time is directly connected to the diffusion coefficient through Eq.~\ref{eq:Diff_tauhop} because the particle must be trapped long enough to perform SFD between hops. However, our TST approach begins to break down for wider channels where we appear to underestimate the height of the barrier needed to predict the hopping time.

% Hard discs transition state scaling
\begin{figure}[ht]
\includegraphics[width=3.5in]{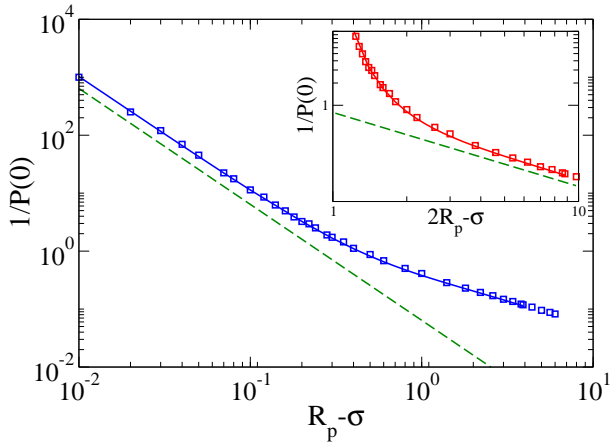}
\caption{Log--Log plot of $1/P(0)$ versus the $R_p-\sigma$ showing the power law behaviour of $\tau_{hop}$ for hard discs as the passing threshold is approached.  The solid line shows the theoretical results obtained from Eq.~\ref{eq:p0dxhd}, the points are the results from our simulation free energy calculations and the dashed line shows the limiting slope of $-2$. Insert: Log--Log plot of $1/P(0)$ versus $2R_p-\sigma$ showing the wide channel scaling law for $\tau_{hop}$ obtained from theory (solid line) and simulation (points). The dashed line highlights the limiting slope of $-1$.}
\label{fig:HD_scaling}
\end{figure}

% Hard discs hopping time.
\begin{figure}[ht]
\includegraphics[width=3.5in]{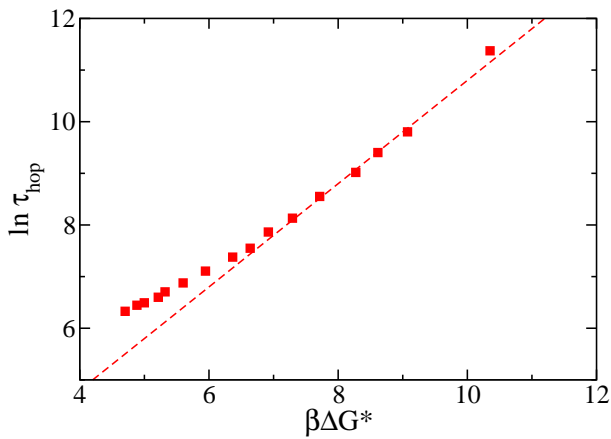}
\caption{ln$\tau_{hop}$ as a function of $\beta\Delta G^*$ for hard discs for a range of different pore radii, $R_p=1.01-1.20$. The points represent the data obtained from simulation with $x_c=0.03$. The dashed line has a slope of 1 and provides a guide to the eye.}
\label{fig:HD_hop}
\end{figure}

%\Discussion
\section{Discussion}
\label{sec:dis}
The phenomenological hopping time approach to understanding diffusion in single file fluids confined within quasi--one dimensional channels is attractive because all the effects that influence particle motion in the long time limit are contained within a single parameter, $\tau_{hop}$, that describes the short time event of a particle hopping past one of its neighbours. The activated nature of the hopping process also makes the calculation of $\tau_{hop}$ amenable to the application of transition state theory, but the challenge is to provide a rigorous approach for the calculation of the probability of being in the transition state, i.e. the height of the hopping free energy barrier, for a nanoscale system that exhibits volume fluctuations. While the isobaric--isothermal ensemble is the appropriate choice of ensemble, a direct application of the bulk thermodynamic results to a system in the nanoscale is problematic due to the system size dependence of the necessary volume scale. 

% key results
We have shown that this problem can be overcome by using the small system isobaric--isothermal ensemble proposed by Corti, where the use of a shell particle to define the volume prevents the over counting of configurations when considering volume fluctuations and eliminates the need for an explicit volume scale. Figures~\ref{fig:2D_HOP_DF} and \ref{fig:HD_hop} highlight the key results of our work and show that $\ln \tau_{hop}$ varies linearly with $\Delta G^*$ with the slope of unity for both systems studied here, over a range of different radii, as predicted by TST. One of the main advantages of our approach is that it only requires the simulation of two fluid particles and the wall interactions, while the direct measurement of the diffusion coefficient and $\tau_{hop}$ involves the simulation of hundreds of particles and requires a long simulation time because $\tau_{hop}$ is slow to converge. The small number of particles also means that the probability density along the reaction coordinate can be sampled by ``brute force" because simulations with a large number of MC cycles can be performed quickly, but the method can also be adapted to include more advanced techniques such as biased umbrella sampling~\cite{Frenkel2002} in cases where the barrier is large.

Choosing the entire partition function of the caged particle as the appropriate reference state to define $\Delta G^*$ is critical to the implementation of our approach. As noted earlier this means that $\Delta G^*$ is necessarily positive because the configuration space of the transition state is smaller than that of the entire system, even for the case of the ideal gas where there is no volume exclusion by the particles as they approach each other. This, combined with the use of the small system $n,p,T$ ensemble, leads to the unexpected prediction that $\tau\sim1/R_p$ which was confirmed through our simulations. 

It is interesting to note that our TST approach breaks down for wider channels in the hard disc model, but works well over all channel radii in the ideal gas model even when the barrier is only a few $kT$. The main difference between the two models is the exclusion interaction of the hard discs and the effect this has on our decision to neglect the interactions between the small system and its surroundings. Setting $U_{\sigma}=0$ is exact for the ideal gas system. In the hard discs model, this allows the two particles in small system to sample configurations in the transition state that would otherwise be ruled out by the exclusion interaction with particles in the surroundings sitting near the boundary. As a result, our method over estimates the partition function of the transition state which leads to a lower predicted barrier. Neglecting these interactions becomes less important as the channel becomes narrower. This assumption also restricts the application of our method to low pressures, where again, interactions between the small system and the surroundings are minimal.

The hopping time is only connected to the diffusion coefficient through Eq.~\ref{eq:Diff_tauhop}, with $\gamma=1/2$, when the particles are caged long enough between hops that they perform single file diffusion during this time, which should occur when the barrier is high. This suggests that our method should be effective in predicting the hopping time and dynamic properties of single file fluids in the crossover regimes and there are a wide range of important practical challenges in nano-fluidic systems, including the separation of mixtures, that can be addressed in the regime where our method is valid and simple enough to provide useful insight. However, Mon and Percus~\cite{Mon:2002p4642} also showed that Eq.~\ref{eq:Diff_tauhop} may be applicable in the wide channel regime, but with a different hopping time exponent. When the channel is wide, the particles can pass each other easily because the barrier is low. The hopping time is then short so that the distance traveled between hops becomes independent of the hopping time, and proportional to the average distance between nearest neighbors in the channel, which leads to an exponent of $\gamma=1$. This has not been tested, but suggests the hopping time approach may still be applicable even when the channel is wide. To make our TST approach work in the wide channel regime, it would be essential to account for the interactions between the small system and the surroundings in our simulation method.

% Conclusions
\section{Conclusions}
\label{sec:con}
In this work, we have shown that the small system isobaric--isothermal ensemble provides a rigorous framework for the development of a transition state theory for the calculation of the hopping time and we have applied it to two simple systems. While the simulation method is currently restricted to the narrow channel, low pressure regime, where interactions between the small system and its surroundings can be neglected, it is simple to implement and is potentially applicable to the study of a wide range of systems.

\acknowledgments
We would like to thank Natural Sciences and Engineering Research Council of Canada (NSERC) and POS Bio-Sciences for financial support. Computing resources and support were provided by Compute Canada and WestGrid.

%\bibliography{SFD_Sheida.bib}

%merlin.mbs aipnum4-1.bst 2010-07-25 4.21a (PWD, AO, DPC) hacked
%Control: key (0)
%Control: author (8) initials jnrlst
%Control: editor formatted (1) identically to author
%Control: production of article title (0) allowed
%Control: page (1) range
%Control: year (1) truncated
%Control: production of eprint (0) enabled
%

\end{document}